\documentclass[seceq,twocolumn]{jpsj2} 
\usepackage{txfonts}

\def\runauthor{Hiroaki \textsc{Kusunose}}

\title{
Influence of Spatial Correlations in Strongly Correlated Electron Systems:\\
Extension to Dynamical Mean Field Approximation
}

\author{
Hiroaki \textsc{Kusunose}\thanks{kusu@cmpt.phys.tohoku.ac.jp}
}

\inst{
Department of Physics, Tohoku University, Sendai 980-8578
}

\abst{
We propose a formalism to take account of the correction of the spatial fluctuations to the local self-energy obtained by the dynamical mean-field approximation.
For this purpose, the approximate dynamical susceptibility in the framework of the iterated perturbation theory is proposed and examined.
Using the formalism, it is demonstrated that the one-particle spectral intensity in the two-dimensional Hubbard model at half-filling exhibits the pseudo-gap behavior in the central coherent quasiparticle peak due to the critical antiferromagnetic fluctuation.
The specific heat is considerably enhanced by the short-range order, which assists a tendency of the Mott localization showing the reduction of the double occupancy.
We briefly discuss a formulation for the superconducting transition temperature in the present approximation.
}

\kword{
dynamical mean field approximation, spin fluctuation, metal-insulator transition, Hubbard model, iterated perturbation theory, anisotropic superconductivity
}

\begin{document}
\maketitle

\section{Introduction}
Dynamics of strongly correlated electron systems has been investigated extensively in context of low-temperature physics in the heavy fermion compounds, the transition metals and some organic materials.
One of the key feature is a duality of electronic states in one-particle density of states (DOS).
Namely, the high-energy incoherent states (so-called the Hubbard bands) have a spatially localized character, while the low-energy states exhibit an itinerant character with strong temperature dependence.
Particle-hole excitations within the upper or lower Hubbard bands constitute the localized spin degrees of freedom, which govern the low-energy spin dynamics.
This low-energy spin fluctuation in the two-particle correlation can have a significant influence on the low-energy itinerant quasiparticle state\cite{Kuramoto90,Ohkawa92,Ohkawa98}.

Various theoretical approaches have been devoted to investigate such an interesting but complicated dynamics.
The representative theories among them may be classified by their nature of the irreducible vertices, i.e., a source of the effective two-body interactions.
They depend on the incoming and the outgoing momenta and energies in general\cite{Abrikosov63}.
The random phase approximation (RPA) approximates the irreducible vertices to bare interactions, while one-particle states are considered in the mean-field level.
Reflecting the RPA-type two-particle fluctuations to the one-particle self-energy in a self-consistent manner, we obtain the fluctuation exchange (FLEX), or the Migdal-Eliashberg approximation\cite{Bickers89,Bickers91}.
This is a conserving approximation in the sense of Baym-Kadanoff\cite{Baym62,Kadanoff62,Brandt70}, but it violates a number of crucial sum-rules\cite{Vilk97}.
On the contrary, the two-particle self-consistent (TPSC) theory respects primarily the two-particle fluctuations and their sum rules rather than the one-particle quantities\cite{Vilk97}.
Following this philosophy, local and instantaneous (contact-type) irreducible vertices are introduced independently in the charge and the spin channels, which are determined so as to satisfy the local-moment sum-rules in each channel.
The one-particle self-energy is then improved by including the effect of the collective modes.
All of the approaches mentioned above is the weak-coupling theory in the sense that they neglect energy dependences in the irreducible vertices.
Therefore, they fail to describe the spin dynamics of the localized spin degrees of freedom in strongly correlated electron systems.
The renormalized energy scales such as a superexchange interaction is never involved in the two-particle correlation.
Nevertheless, it is worth noticing that the weak-coupling theories using the contact-type irreducible vertices qualitatively account for the characteristic momentum dependences inherent from the electronic band structures\cite{Kontani99}.

On the other hand, the dynamical mean-field theory (DMFT) has been proven successfully to describe the dual nature of the electronic states in the energy space, e.g., the opening of the Mott gap at half-filling for sufficiently strong Coulomb repulsion\cite{Georges96,Georges04}.
If we remind the fact that the self-energy is expressed in terms of the vertices\cite{note1},
the proper energy dependences in the irreducible vertices have crucial importance in describing the duality of the one-particle electronic states as well as the renormalized energy scale in the two-particle correlations.
The one-particle self-energy in DMFT does not contain any contributions from the nonlocal correlations.
Hence, extensive studies such as cluster extensions of DMFT have been carried out to take account of short-range correlations\cite{Schiller95,Hettler98,Maier05,Kyung06}.

In this paper, we propose an alternative extension to DMFT by taking account of the corrections of the spatial fluctuations in the self-energy.
Sadovskii and co-workers have proposed an introduction of the effective length scale of the paramagnon excitations for this purpose\cite{Sadovskii05,Kuchinskii05}.
We assume local but energy-dependent irreducible vertices to construct the nonlocal two-particle correlations.
This is a minimum requirement to describe the two-particle correlations both with renormalized energy scale and band-specific features in the momentum space.
The resultant one-particle states with the nonlocal self-energy can represent the duality in the energy space with a proper momentum dependence.
A similar approach named as the dynamical vertex approximation (D$\Gamma$A) has recently been developed on the basis of the parquet equation\cite{Toschi06}.
Specifically, we extract the local irreducible vertices from the result of DMFT, and construct the nonlocal dynamical susceptibilities.
Then the nonlocal corrections of the spatial correlations are taken into account to improve the DMFT self-energy analogous to the TPSC theory.
For this purpose, we develop an approximate construction for the local dynamical susceptibility based on the iterated perturbation theory (IPT)\cite{Georges92,Rozenberg94,Kajueter96,Kajueter96a,Potthoff97}.
Then, we discuss the influence of the spatial correlations in the one-particle spectral intensity for the symmetric 2D Hubbard model at half-filling.

The paper is organized as follows.
In \S2, we establish the formulation of the extension to DMFT.
The subsequent section is devoted to explain the construction of the dynamical spin susceptibility based on IPT.
The validity of the approximate spin susceptibility and the $U$-$T$ phase diagram are discussed.
The main result of the paper is given in \S4.
The influence of the spatial correlations in the one-particle spectral intensity is mainly discussed.
The summary and discussion are presented in the last section.
The outline of the IPT scheme extensively used in this paper is summarized in Appendix A.
A formulation for the superconducting transition temperature based on the present approximation is briefly discussed in Appendix B.

\section{Extension to DMFT}
In this section, we describe the formulation to take into account the correction of the spatial correlations to the DMFT self-energy.
To be specific, we consider the Hubbard model on the two-dimensional square lattice,
\begin{equation}
H=\sum_{\mib{k},\sigma}\epsilon_{\mib{k}}c^\dagger_{\mib{k}\sigma}c_{\mib{k}\sigma}
+U\sum_i n_{i\uparrow}n_{i\downarrow},
\end{equation}
where $c^\dagger_{\mib{k}\sigma}$ is a creation operator of conduction electron with the wave-vector $\mib{k}=(k_x,k_y)$ and the spin $\sigma$. The electron dispersion is given by $\epsilon_{\mib{k}}=-2t(\cos k_xa+\cos k_ya)$.
The overlap integral $t$ and the lattice constant  $a$ are set by $t=a=1$.
$n_{i\sigma}=c^\dagger_{i\sigma}c_{i\sigma}$ is the density operator at the site $i$.
For notational simplicity, we use the (2+1)-vector representation, $k=(\mib{k},i\omega_n)$, if necessary.
The 3-vector summation, $\sum_k$, represents $(T/N)\sum_{\mib{k}}\sum_n$, $N$ being the number of sites.
We also use the notational conventions, (i) the letters $k$ and $\omega_n$ ($q$ and $\epsilon_m$) are used to represent the fermionic (bosonic) Matsubara frequency, and (ii) a quantity computed by DMFT is indicated by the superscript, L.
We consider only the symmetric case with the electron density, $n=1$, throughout this paper.

\begin{figure}
\includegraphics[width=8.5cm]{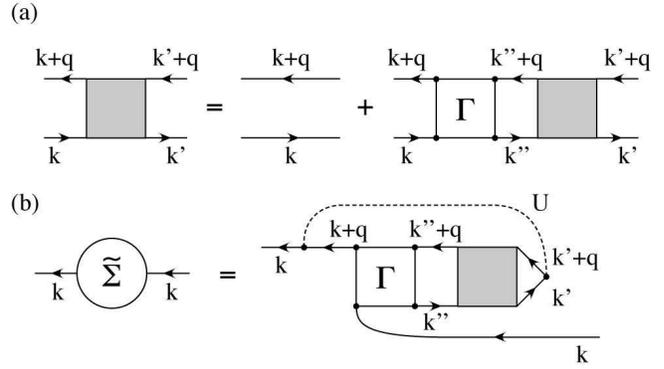}
\caption{The Feynman diagrams for (a) the Bethe-Salpeter equation, (\ref{bethe-salpeter}), (b) the self-energy in terms of susceptibilities and irreducible vertices, (\ref{self-energy}).}
\label{diagram}
\end{figure}

We begin with two exact relations.
One is the Bethe-Salpeter (BS) equation in the charge and the spin channels\cite{Abrikosov63,Kadanoff62,Vilk97},
\begin{subequations}\label{bethe-salpeter}
\begin{align}
&\chi_{\rm ch}(k,k';q)=\chi_0(k;q)\biggl[
\delta_{k,k'}
\notag\\&\quad\quad\quad\quad
-\sum_{k''}\Gamma_{\rm ch}(k,k'';q)\chi_{\rm ch}(k'',k';q)\biggr],\\
&\chi_{\rm sp}(k,k';q)=\chi_0(k;q)\biggl[
\delta_{k,k'}
\notag\\&\quad\quad\quad\quad
+\sum_{k''}\Gamma_{\rm sp}(k,k'';q)\chi_{\rm sp}(k'',k';q)\biggr],
\end{align}
\end{subequations}
where $\chi_0(k;q)=-G(k)G(k+q)$ is the lowest-order two-particle Green's function.
$\chi_{\rm ch}(k,k';q)$ and $\Gamma_{\rm ch}(k,k';q)$ are the two-particle Green's function and the irreducible vertex in the charge channel.
The similar notation is applied for the longitudinal spin channel.
The sign of the irreducible vertices are chosen so as to become the positive $U$ in the lowest order, i.e. $\Gamma_{\rm ch}(k,k';q)=\Gamma_{\rm sp}(k,k';q)=U$ for RPA or FLEX approximations.
Note that $\Gamma_{\rm ch}(k,k';q)=U_{\rm ch}$ and $\Gamma_{\rm sp}(k,k';q)=U_{\rm sp}$ are used in TPSC.
The ordinary susceptibility is obtained by the summations,
\begin{equation}
\chi_{\alpha}(q)=\sum_{k,k'}\chi_\alpha(k,k';q),
\quad
(\alpha=\text{ch, sp}).
\end{equation}
Note that we have defined the (spin) density operator as $\rho_{\rm ch (sp)}=(c^\dagger_\uparrow c_\uparrow\pm c^\dagger_\downarrow c_\downarrow)/\sqrt{2}$ in the susceptibilities.
The Feynman diagram for the BS equation is shown in Fig.~\ref{diagram}(a).

The Dyson's equation for the one-particle Green's function is given by
\begin{equation}\label{dyson}
G^{-1}(k)=i\omega_n+\tilde{\mu}-\epsilon_{\mib{k}}-\tilde{\Sigma}(k),
\end{equation}
where $\tilde{\Sigma}(k)$ is the self-energy in which the Hartree contribution is subtracted.
Namely, the chemical potential, $\tilde{\mu}$, contains the Hartree contribution as $\tilde{\mu}=\mu-Un/2$ where $n=2\sum_{k}G(k)e^{i\omega_n0_+}$.
We omit spin indices since we will discuss only the paramagnetic state.
The self-energy can be expressed in terms of the charge and the (longitudinal) spin susceptibilities\cite{Abrikosov63,Kadanoff62,Vilk97} as
\begin{multline}
\tilde{\Sigma}(k)=\frac{U}{2}\sum_q\sum_{k',k''}\biggl[
\Gamma_{\rm ch}(k,k'';q)\chi_{\rm ch}(k'',k';q)\\
+\Gamma_{\rm sp}(k,k'';q)\chi_{\rm sp}(k'',k';q)\biggr]G(k+q).
\label{self-energy}
\end{multline}
This is the second exact relation, and its Feynman diagram is shown in Fig.~\ref{diagram}(b).
It should be emphasized that we can also write down the self-energy in terms of alternative fluctuation, such as superconducting (pairing) one\cite{Allen01}.
Here, we have chosen the expression of the charge and the spin fluctuations, since the latter fluctuation plays a dominant role in strongly repulsive $U$ regions.
For negative $U$ problem, the expression of the pairing fluctuation is appropriate\cite{Allen01}.

In order to solve the above two equations, let us introduce two approximations as follows.
As was mentioned in the introduction, the energy dependence in the irreducible vertices is significantly important in strongly correlated electron systems.
Therefore, as the simplest approximation, we retain only the external energy dependence, $\epsilon_m$, in the irreducible vertices,
\begin{equation}
\Gamma_\alpha(k,k';q)\sim \Gamma_\alpha(\epsilon_m).
\end{equation}
Then, the BS equations, (\ref{bethe-salpeter}), can be solved formally as
\begin{subequations}\label{bs-approx}
\begin{align}
&\chi_{\rm ch}^{-1}(q)=\chi_0^{-1}(q)+\Gamma_{\rm ch}(\epsilon_m),
\\
&\chi_{\rm sp}^{-1}(q)=\chi_0^{-1}(q)-\Gamma_{\rm sp}(\epsilon_m).
\end{align}
\end{subequations}
Note that the momentum dependence is considered only through $\chi_0(q)$.
Within the same approximation, we can divide the contributions from the fluctuations in eq.~(\ref{self-energy}) into the local and the nonlocal ones,
\begin{multline}\label{se-approx}
\tilde{\Sigma}(k)=\tilde{\Sigma}(\omega_n)+\frac{U}{2}\sum_q\biggl[
\Gamma_{\rm ch}(\epsilon_m)\Delta\chi_{\rm ch}(q)\\
+\Gamma_{\rm sp}(\epsilon_m)\Delta\chi_{\rm sp}(q)\biggr]G(k+q),
\end{multline}
where $\Delta\chi_\alpha(q)=\chi_\alpha(q)-\chi_\alpha(\epsilon_m)$ describes the nonlocal contribution.
The local self-energy is defined as
\begin{multline}
\tilde{\Sigma}(\omega_n)=\frac{U}{2}T\sum_{\epsilon_m}\biggl[
\Gamma_{\rm ch}(\epsilon_m)\chi_{\rm ch}(\epsilon_m)
\\
+\Gamma_{\rm sp}(\epsilon_m)\chi_{\rm sp}(\epsilon_m)\biggr]G(\omega_n+\epsilon_m),
\end{multline}
where the local Green's function is given by $G(\omega_n)=(1/N)\sum_{\mib{k}}G(k)$.

Next, we introduce the second approximation.
The local quantities in eqs.~(\ref{bs-approx}) and (\ref{se-approx}) are replaced with those obtained by DMFT,
\begin{equation}
\tilde{\Sigma}(\omega_n)\sim \tilde{\Sigma}^{\rm L}(\omega_n),
\end{equation}
\begin{equation}
\chi_\alpha(\epsilon_m)\sim \chi_\alpha^{\rm L}(\epsilon_m),
\end{equation}
and the local vertices are calculated from the relations similar to eqs.~(\ref{bs-approx}),
\begin{subequations}\label{vertices}
\begin{align}
&+\Gamma_{\rm ch}(\epsilon_m)\sim 1/\chi^{\rm L}_{\rm ch}(\epsilon_m)-1/\chi^{\rm L}_0(\epsilon_m),\\
&-\Gamma_{\rm sp}(\epsilon_m)\sim 1/\chi^{\rm L}_{\rm sp}(\epsilon_m)-1/\chi^{\rm L}_0(\epsilon_m),
\end{align}
\end{subequations}
where the local irreducible susceptibility is given by $\chi^{\rm L}_0(\epsilon_m)=-T\sum_{\omega_n}G^{\rm L}(\omega_n)G^{\rm L}(\omega_n+\epsilon_m)$.
With these approximations, (\ref{se-approx}) reads the equation to improve the local DMFT self-energy by including the corrections of the spatial correlations.
Then, the updated Green's function is obtained by the Dyson's equation with the improved self-energy.
The chemical potential $\tilde{\mu}$ is determined such that $2\sum_k G(k)e^{i\omega_n0_+}$ coincides with the given electron density\cite{Vilk97}.
In this paper, we always use $\tilde{\mu}=0$ because of the symmetric condition.

\section{Dynamical Susceptibility Based on IPT}
In the previous section, we propose the formulation to improve the local self-energy by including the correction of the nonlocal bosonic fluctuations.
To perform a practical calculation, the dynamical susceptibilities as well as the local self-energy are required within DMFT.
In practice, however, efficient algorithms to compute the dynamical susceptibilities at finite temperature are very limited.
In fact the quantum monte carlo (QMC) simulation based on the Hirsch-Fye algorithm is only such an efficient algorithm at moment\cite{Georges96,Jarrell92,Hirsch86}.
In this section, we first develop an approximate construction of the dynamical susceptibility based on IPT.
Then, we examine the validity of the approximate dynamical susceptibility.

\subsection{Formulation of dynamical susceptibilities}
Within the spirit of DMFT, we have the local version of the relations\cite{Georges96,Jarrell92,Zlatic90}, eqs.~(\ref{bethe-salpeter}) and (\ref{self-energy}),
\begin{subequations}
\begin{align}
&\chi_{\rm ch}^{\rm L}(\omega_n,\omega_n';\epsilon_m)=
\chi_0^{\rm L}(\omega_n;\epsilon_m)\biggl[
T^{-1}\delta_{\omega_n,\omega_n'}
\notag\\&\quad\quad\quad
-T\sum_{\omega_n''}\Gamma^{\rm L}_{\rm ch}(\omega_n,\omega_n'';\epsilon_m)\chi^{\rm L}_{\rm ch}(\omega_n'',\omega_n';\epsilon_m)\biggr],\\
&\chi^{\rm L}_{\rm sp}(\omega_n,\omega_n';\epsilon_m)=
\chi_0^{\rm L}(\omega_n;\epsilon_m)\biggl[
T^{-1}\delta_{\omega_n,\omega_n'}
\notag\\&\quad\quad\quad
+T\sum_{\omega_n''}\Gamma^{\rm L}_{\rm sp}(\omega_n,\omega_n'';\epsilon_m)\chi^{\rm L}_{\rm sp}(\omega_n'',\omega_n';\epsilon_m)\biggr],
\end{align}
\end{subequations}
and
\begin{multline}\label{local-se}
\tilde{\Sigma}^{\rm L}(\omega_n)=\frac{U}{2}T^3\sum_{\epsilon_m}\sum_{\omega_n',\omega_n''}\biggl[
\Gamma^{\rm L}_{\rm ch}(\omega_n,\omega_n'';\epsilon_m)\chi^{\rm L}_{\rm ch}(\omega_n'',\omega_n';\epsilon_m)\\
+\Gamma^{\rm L}_{\rm sp}(\omega_n,\omega_n'';\epsilon_m)\chi^{\rm L}_{\rm sp}(\omega_n'',\omega_n';\epsilon_m)\biggr]G^{\rm L}(\omega_n+\epsilon_m).
\end{multline}
Eliminating the irreducible vertices from these relations, we obtain
\begin{multline}\label{se-local}
\tilde{\Sigma}^{\rm L}(\omega_n)=\frac{U}{2}T\sum_{\epsilon_m}\biggl[
\chi^{\rm L}_{\rm sp}(\omega_n;\epsilon_m)-\chi^{\rm L}_{\rm ch}(\omega_n;\epsilon_m)\biggr]\\
\times G^{\rm L}(\omega_n+\epsilon_m)/\chi_0^{\rm L}(\omega_n;\epsilon_m),
\end{multline}
where $\chi_\alpha^{\rm L}(\omega_n;\epsilon_m)=T\sum_{\omega_n'}\chi_\alpha^{\rm L}(\omega_n,\omega_n';\epsilon_m)$.
This expression implies that the local self-energy contains information of the two-particle susceptibility (it is nothing but the spin antisymmetrized contribution).

Meanwhile, the self-energy in IPT (see Appendix A for detail) has the form,
\begin{equation}
\tilde{\Sigma}^{\rm L}(\omega_n)=\frac{U}{2}T\sum_{\epsilon_m}K(\omega_n)\bar{\chi}_0(\epsilon_m)\tilde{\cal G}(\omega_n+\epsilon_m),
\end{equation}
with $K(\omega_n)=2UA/[1-B\Sigma^{(2)}(\omega_n)]$ and $\bar{\chi}_0(\epsilon_m)=-T\sum_{\omega_n}\tilde{\cal G}(\omega_n)\tilde{\cal G}(\omega_n+\epsilon_m)$, where $\Sigma^{(2)}(\omega_n)$ being the second-order self-energy with respect to the cavity Green's function, $\tilde{\cal G}(\omega_n)$.
Comparing this expression with eq.~(\ref{se-local}), we obtain
\begin{multline}
\chi^{\rm L}_{\rm sp}(\epsilon_m)-\chi^{\rm L}_{\rm ch}(\epsilon_m)=-\bar{\chi}_0(\epsilon_m)
T\sum_{\omega_n}K(\omega_n)G^{\rm L}(\omega_n)\\
\times[\tilde{\cal G}(\omega_n+\epsilon_m)+\tilde{\cal G}(\omega_n-\epsilon_m)]/2.
\label{ipt-susceptibility}
\end{multline}
Note that $\chi^{\rm L}_{\rm sp}(\epsilon_m)$ is written in the symmetrized form with respect to $\epsilon_m\leftrightarrow -\epsilon_m$.
This expression can be calculated by using only the one-particle quantities in IPT.

In order to separate the charge and the spin contributions, we further approximate $\Gamma^{\rm L}_{\rm ch}(\epsilon_m)\sim U_{\rm ch}$, which will be determined so as to satisfy the sum-rule\cite{Vilk97},
\begin{equation}
T\sum_{\epsilon_m}\chi^{\rm L}_{\rm ch}(\epsilon_m)=\frac{1}{2}\left[ n(1-n) + 2d \right],
\end{equation}
with $\chi^{\rm L}_{\rm ch}(\epsilon_m)=\chi^{\rm L}_0(\epsilon_m)/[1+U_{\rm ch}\chi^{\rm L}_0(\epsilon_m)]$.
The double occupancy, $d=\langle n_{i\uparrow}n_{i\downarrow}\rangle$, is computed by IPT, i.e., $d=(n/2)^2+(T/U)\sum_{\omega_n}\tilde{\Sigma}^{\rm L}(\omega_n)G^{\rm L}(\omega_n)$.
Then, using eq.~(\ref{ipt-susceptibility}) and $\chi^{\rm L}_{\rm ch}(\epsilon_m)$ with $U_{\rm ch}$, we can compute the approximate local spin susceptibility within the IPT scheme.
The charge vertex $U_{\rm ch}$ is turned out to be rapidly increase function of $U$, and hence the charge susceptibility is negligible in strong $U$ region.

\subsection{Validity of approximate spin susceptibility}
\begin{figure}
\includegraphics[width=8.5cm]{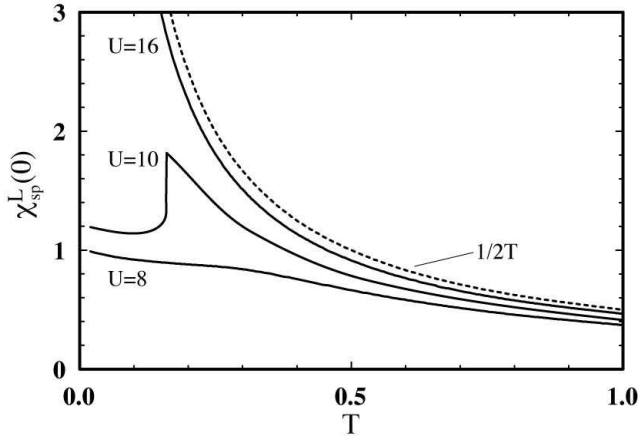}
\caption{The $T$ dependence of the static local spin susceptibility for $U=8$, $10$ and $16$. The dotted line represents the Curie law in the atomic limit.}
\label{chiL-t}
\end{figure}
Let us examine the validity of the approximate spin susceptibility as follows.
We have used the tight-binding DOS, $\rho_0(\epsilon)$, of the two-dimensional square lattice with the 2048 Matsubara frequencies in solving the DMFT equations.
For the summations of the convolution type, the use of the Fast Fourier Transform (FFT) algorithm is quite efficient\cite{Bickers91,Deisz02}.
Retarded quantities are obtained by the numerical analytic continuation to the upper half plane, i.e., $i\omega_n\to\omega+i\delta$ $(i\epsilon_m\to\epsilon+i\delta)$, using the Pad\'e approximant\cite{Vidberg77,NR02}.

\begin{figure}
\includegraphics[width=8.5cm]{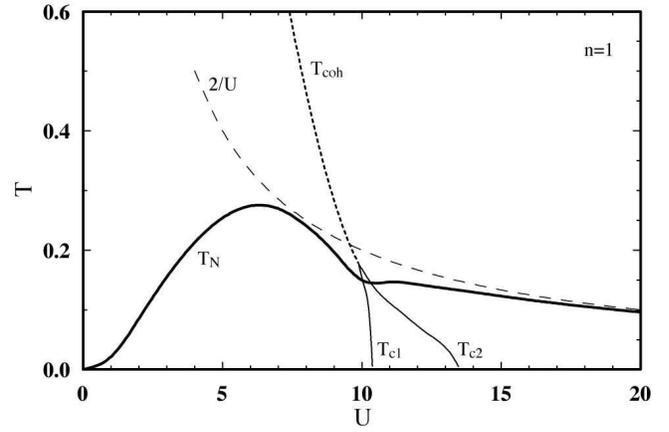}
\caption{The $U$-$T$ phase diagram for half-filling. $T_{c1}$ and $T_{c2}$ are the metal-insulator transition temperatures determined by DMFT. $T_{\rm coh}$ is the coherent temperature, below which the coherent state develops near the chemical potential. The N\'eel temperature $T_{\rm N}$ is determined by eq.~(\ref{bs-approx}b).}
\label{phase}
\end{figure}

The $T$ dependence of the static local spin susceptibility for $U=8$, $10$ and $16$ is shown in Fig.~\ref{chiL-t}.
The dotted line represents the Curie law in the atomic limit, ($\chi_{\rm sp}^{\rm L}(\epsilon_m=0)=1/2T$).
The $T$ dependence of $\chi_{\rm sp}^{\rm L}(0)$ approaches the Curie law as $U$ increases.
It is important to note that a simple irreducible vertex based on a weak-coupling perturbation theory cannot reproduce the behaviors in the atomic limit (the relevant energy scale, $J$, is {\it inversely} proportional to $U$).
The discontinuous jump at $T\sim 0.18$ for $U=10$ is due to the metal-insulator transition (see the $U$-$T$ phase diagram in Fig.~\ref{phase}).
Such a singular behavior is confirmed by the QMC calculation\cite{Georges96}.

\begin{figure}
\includegraphics[width=8.5cm]{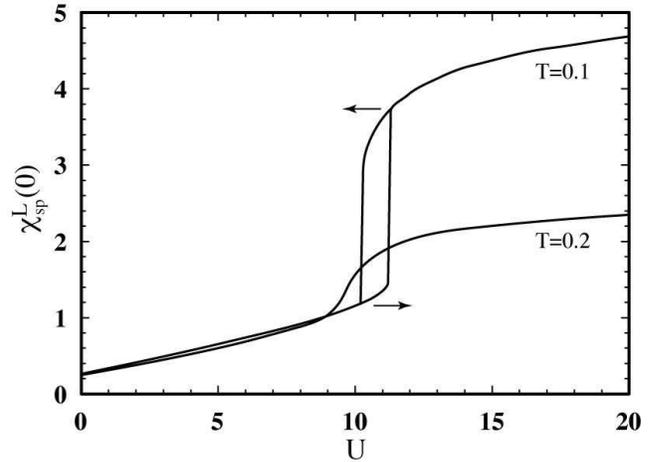}
\caption{The $U$ dependence of the static local spin susceptibility for $T=0.1$ and $0.2$. The arrows indicate the hysteresis loop.}
\label{chiL-u}
\end{figure}
The $U$ dependence of $\chi_{\rm sp}^{\rm L}(0)$ for $T=0.1$ and $0.2$ is shown in Fig.~\ref{chiL-u}.
$\chi_{\rm sp}^{\rm L}(0)$ for $T=0.2$ rapidly increases around the metal-insulator crossover region.
Two discontinuous jumps for $T=0.1$ are due to the first-order metal-insulator transition.
The hysteresis loop of the first-order transition is indicated by the arrows.
These behaviors are also confirmed by the QMC calculation\cite{Georges96}.

\begin{figure}
\includegraphics[width=8.5cm]{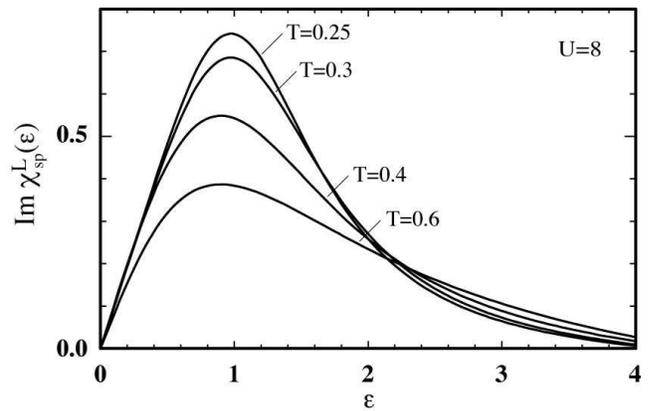}
\caption{The spectral intensity of the local spin susceptibility for $U=8$ and $T=0.6$, $0.4$, $0.3$ and $0.25$.}
\label{chiL-e}
\end{figure}

The spectral intensity of $\chi_{\rm sp}^{\rm L}(\epsilon)$ for $U=8$ is shown in Fig.~\ref{chiL-e}.
The low-energy weight of the spectral intensity develops as $T$ decreases.
Notice that the peak of the intensity is located at the order of the superexchange interaction, $J\sim t^2/U$, irrespective of the presence of the quasiparticle coherent states at $\omega=0$ (see, Figs.\ref{phase} and \ref{dos-u8}b).
This observation supports the assertion emphasized by Vilk and Tremblay\cite{Vilk97} that the bosonic collective modes are weakly dependent on the precise form of the single-particle excitations.
On the other hand, the details of the single-particle self-energy can be strongly influenced by scattering from collective (paramagnon) modes.
Hence, we first compute the two-particle correlations with the DMFT Green's functions, and then we improve on the self-energy by including the effect of collective modes as shown later.

\begin{figure}
\includegraphics[width=8.5cm]{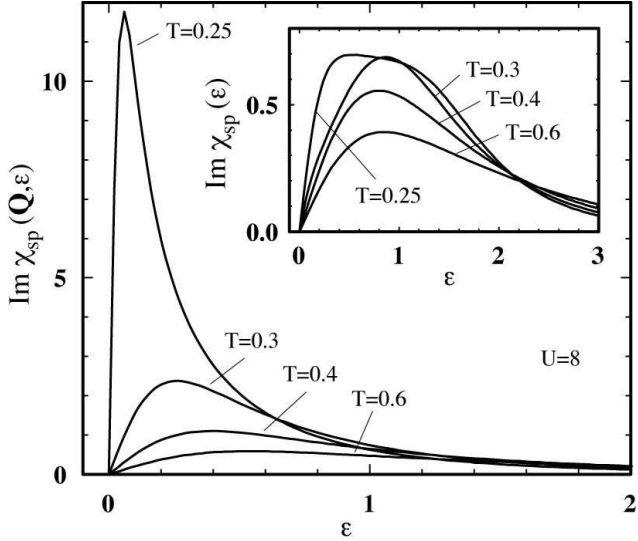}
\caption{The spectral intensity of the spin susceptibility at $\mib{Q}=(\pi,\pi)$ for $U=8$ and $T=0.6$, $0.4$, $0.3$ and $0.25$. The inset shows the spectral intensity of the local spin susceptibility.}
\label{chiQ}
\end{figure}

Using eqs.~(\ref{vertices}) and (\ref{bs-approx}b), the momentum dependence of the spin susceptibility is evaluated.
Figure \ref{chiQ} shows the spectral intensity of $\chi_{\rm sp}(\mib{Q},\epsilon)$ for $U=8$ at $\mib{Q}=(\pi,\pi)$.
Because of the perfect nesting of the Fermi surface, the low-energy weight of the staggered component is strongly enhanced as $T$ approaches the antiferromagnetic (AF) instability, $T_{\rm N}$ (see Fig.~\ref{phase}).
The inset shows the average of ${\rm Im}\,\chi_{\rm sp}(\mib{q},\epsilon)$ over the $\mib{q}$ space.
As compared with Fig.~\ref{chiL-e}, the slope of the local spectral intensity at $\epsilon=0$ becomes much steeper toward the AF instability due to the critical divergence of the staggered spin susceptibility.
It results in the strong enhancement of the NMR relaxation rate, $1/T_1T=\lim_{\epsilon\to0}{\rm Im}\,\chi_{\rm sp}(\epsilon)/\epsilon$.
This critical spin fluctuation simultaneously causes the pseudo-gap behavior in the one-particle DOS as will be discussed later.
Thus, the pseudo-gap behavior due to the spin fluctuation should always be accompanied by the enhancement of $1/T_1T$ rather than a spin-gap behavior\cite{Yasuoka89}.

\begin{figure}
\includegraphics[width=8.5cm]{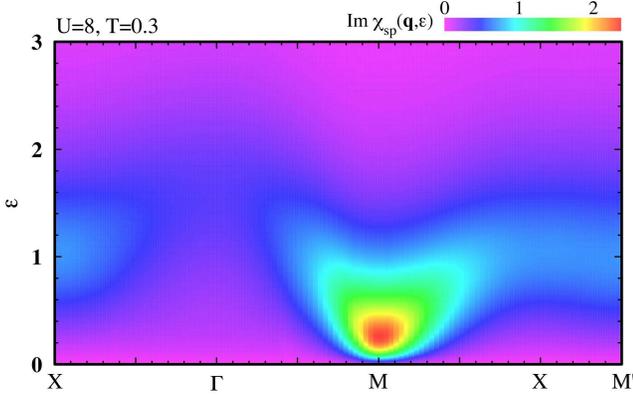}
\caption{The contour plot of the spectral intensity of the spin susceptibility along the high-symmetry lines for $U=8$ and $T=0.3$.}
\label{cqe}
\end{figure}

The contour plot of the spectral intensity of $\chi_{\rm sp}(\mib{q},\epsilon)$ along the high-symmetry lines for $U=8$ and $T=0.3$ is shown in Fig.~\ref{cqe}.
The high-symmetry points are defined as $\Gamma=(0,0)$, ${\rm X}=(\pi,0)$, ${\rm M}=(\pi,\pi)$ and ${\rm M}'=(\pi/2,\pi/2)$.
It exhibits the precursor of the collective spin-wave-like dispersion of the energy-scale $J$ with the highest intensity at $\mib{Q}$.
The spin fluctuation causes a significant modification on the DMFT self-energy, as will be shown in the next section.

\begin{figure}
\includegraphics[width=8.5cm]{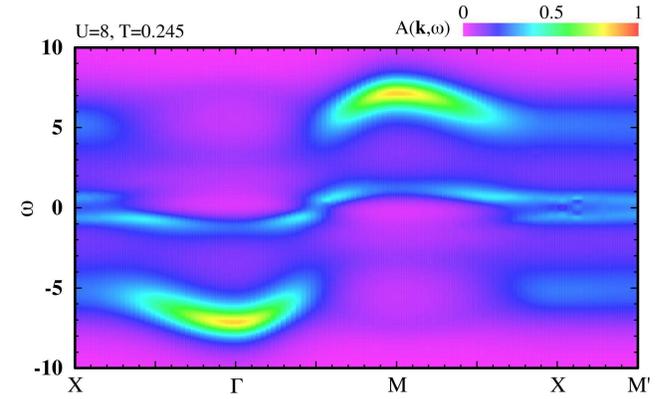}
\caption{The contour plot of the spectral intensity $A(\mib{k},\omega)$ along the high-symmetry lines of the Brillouin zone for $U=8$ and $T=0.245$. The pseudo-gap behavior is observed at the X-M' line.}
\label{akw}
\end{figure}

Now, we evaluate the N\'eel temperature, $T_{\rm N}$ by the condition, $\chi_{\rm sp}^{-1}(\mib{Q},0)=10^{-3}$.
Note that due to the approximate spin vertex, the critical behavior has the mean-field character with finite $T_{\rm N}$ even in two dimension.
The $U$-$T$ phase diagram is shown in Fig.~\ref{phase}.
$T_{c1}$ and $T_{c2}$ are the first-order metal-insulator transition lines determined by IPT.
We define the coherent temperature, $T_{\rm coh}$, at which the central quasiparticle peak begins to appear in the DMFT DOS, $A(\omega)=-{\rm Im}\,G^{\rm L}(\omega)$.
With increase of $U$, $T_{\rm N}$ first increases, showing the maximum at $U\sim 7$, and then decreases followed by $2/U$.
The suppression of $T_{\rm N}$ around $U=10$ is ascribed to the singular behavior of the static local susceptibility, $\chi_{\rm sp}^{\rm L}(0)$, due to the metal-insulator transition (see Figs.~\ref{chiL-t} and \ref{chiL-u}).
In other words, the drop of $\chi_{\rm sp}^{\rm L}(0)$ leads to the suppression of the local spin vertex via eq.~(\ref{vertices}).
In comparison with the result of the QMC\cite{Rozenberg94,Georges96} with the full local vertex, $\Gamma_{\rm sp}^{\rm L}(\omega_n,\omega_n';\epsilon_m)$, the present $T_{\rm N}$ is overestimated in weak-coupling region, and underestimated in strong-coupling region.
The source of the discrepancy mainly comes from the simplification of the spin vertex rather than the IPT approximation itself.
We have checked that the $T_N(U)$ curve becomes similar to the result of the QMC, if we include the $\omega_n$ dependence in the spin vertex (neglecting off-diagonal elements of $\omega_n$ and $\omega_n'$).
However, in strong $U$ region, the actual $T_{\rm N}$ is also expected to be lower than that of DMFT with QMC, since DMFT completely neglects the spatial correlations.
It is also curious to see how the nonlocal spin fluctuation affects the developed coherent quasiparticle state.
This question will be answered in the next section.

\section{Influence of Spatial Correlation}
In this section we discuss the influence of the spatial correlation to the local self-energy based on eq.~(\ref{se-approx}), in which local quantities are evaluated by DMFT (IPT).
We have used $64\times 64$ discretized meshes in the 1st Brillouin zone and 2048 Matsubara frequencies in the computation.

Figure \ref{akw} shows the contour plot of the spectral intensity for $U=8$ and $T=0.245$ along the high-symmetry lines of the Brillouin zone.
The overall structure of the spectral intensity is roughly the same as that obtained by eq.~(\ref{dyson}) with the DMFT self-energy (not shown).
In contrast to the DMFT intensity, the pseudo-gap behaviors at $\omega=0$ appears in the coherent quasiparticle state due to the strong AF spin fluctuation (see, the X-M' line)\cite{Preuss95,Moukouri01,Kyung03,Kyung04,Markiewicz04,Senechal04,Tremblay06}.
Note that there exist the hot spots over the entire Fermi surface, since the magnetic Brillouin zone coincides with the Fermi surface.
The pseudo-gap phenomena due to the AF spin fluctuations are also discussed in the framework of the cluster DMFT\cite{Maier05}.

\begin{figure}
\includegraphics[width=8.5cm]{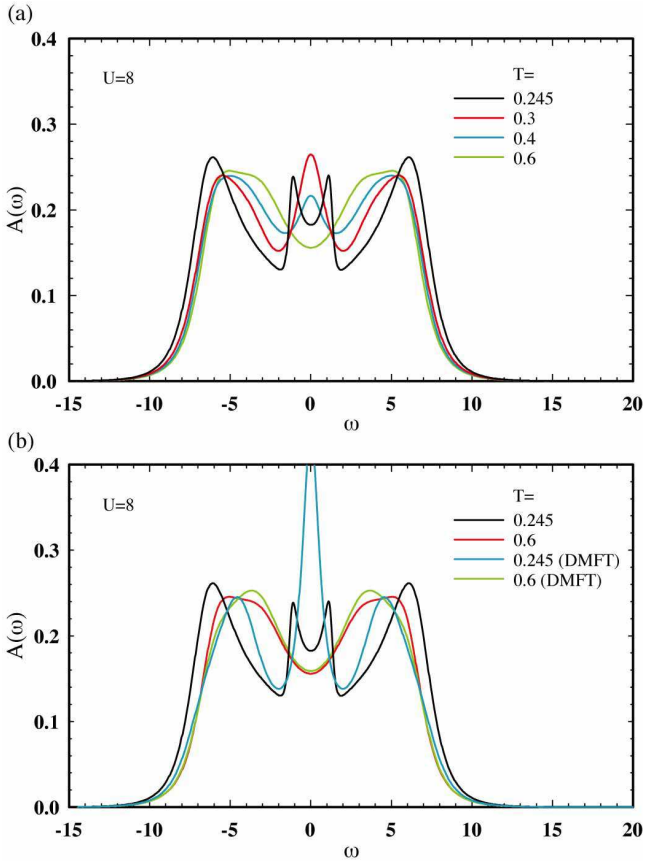}
\caption{The density of states for $U=8$. (a) temperature dependences, (b) comparison with the DMFT results.}
\label{dos-u8}
\end{figure}

To see how the pseudo gap develops with decrease of $T$, the total DOS, $A(\omega)=-(1/N)\sum_{\mib{k}}{\rm Im}G(k)$, for $U=8$ is shown in Fig.~\ref{dos-u8}.
Figure~\ref{dos-u8}(a) exhibits first the development of the coherent quasiparticle peak at $\omega=0$.
As $T$ approaches $T_{\rm N}\doteqdot 0.241$, $A(\omega)$ yields the pseudo-gap behavior, and simultaneously the upper and the lower Hubbard bands become more pronounced.
The comparison with the DMFT DOS is shown in Fig.~\ref{dos-u8}(b).
For the temperature much above $T_{\rm N}$, the inclusion of the spatial correlations affects little on $A(\omega)$, i.e., the spectral weight is slightly shifted to high-energy side.
On the other hand, the quasiparticle peak is significantly suppressed near $T_{\rm N}$ yielding the pseudo-gap behavior, and considerable amount of the weight in Hubbard bands is transferred to high-energy side.
This spectral weight transfer brings the electronic state close to that in the AF ordered phase\cite{Kusunose03,Senechal05}.

\begin{figure}
\includegraphics[width=8.5cm]{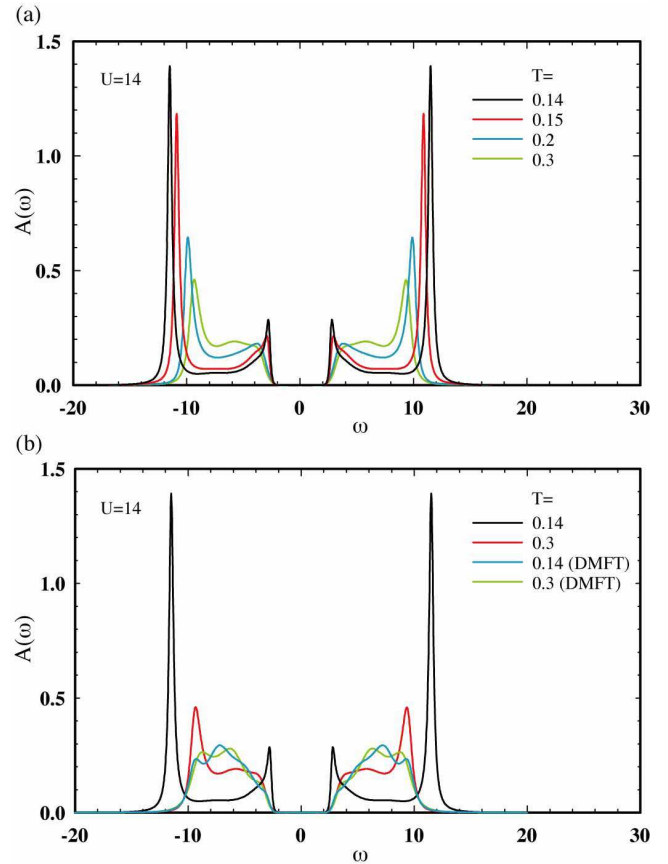}
\caption{The density of states for $U=14$. (a) temperature dependences, (b) comparison with the DMFT results.}
\label{dos-u14}
\end{figure}

The similar plot in the insulating case, $U=14$ is shown in Fig.~\ref{dos-u14}.
In contrast to the results in the metallic case, the low-energy states near the insulating gap edge become more pronounced as $T$ approaches $T_{\rm N}\doteqdot 0.130$.
As similar to the metallic case, the weight of the Hubbard bands is shifted to high-energy side.
However, its intensity is rather sharp as compared with that in the metallic case.
In the present calculation we include only the {\it single} exchange of the spin fluctuations.
If we take account of multiple scattering from the spin fluctuations, the sharp incoherent peak would become much broader.
Note also that the resultant spectral intensity near $T_{\rm N}$ has a close resemblance to that in the AF ordered phase.

\begin{figure}
\includegraphics[width=8.5cm]{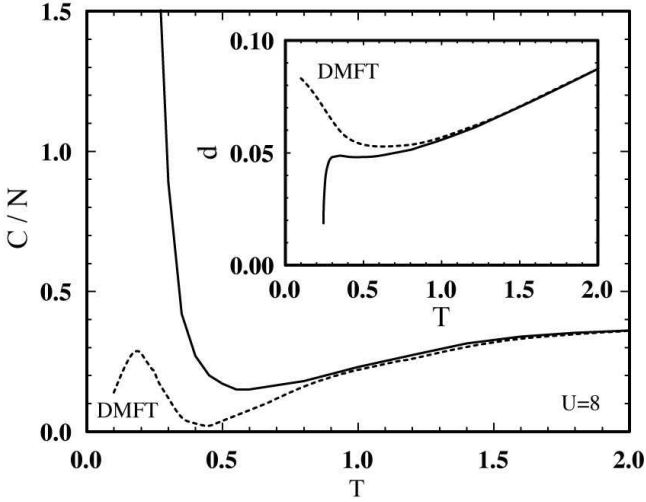}
\caption{The $T$ dependences of the specific heat for $U=8$. The inset shows the $T$ dependence of the double occupancy. The results of DMFT is shown for comparison.}
\label{c-d}
\end{figure}

Finally, we discuss the $T$ dependence of the specific heat, which is calculated by the numerical differentiation of the internal energy,
\begin{equation}
\frac{E}{N}=2\sum_k \epsilon_{\mib{k}}G(k)e^{i\omega_n0_+} + U d.
\end{equation}
Here the double occupancy, $d$ is given by
\begin{equation}
d=\left(\frac{n}{2}\right)^2 + \frac{1}{U}\sum_k \tilde{\Sigma}(k)G(k).
\end{equation}
The $T$ dependence of the specific heat for $U=8$ is shown in Fig.~\ref{c-d}.
The results by IPT are also shown for comparison.
Owing to the short-range AF order, the specific heat is strongly enhanced toward $T_{\rm N}$.
As shown in the inset, the double occupancy rapidly decreases toward the AF instability\cite{Moukouri01} in contrast to the increasing behavior in IPT.
It should be noted that the strong enhancement in the specific heat near $T_{\rm N}$ mainly comes from the reduction of $d$ rather than that of the kinetic energy.
In other words, the AF correlation enhances the tendency of the Mott localization.
The QMC simulation using the $6\times6$ square lattice shows the $T$ dependence of the specific heat similar to that of the DMFT\cite{Paiva01}.
This implies the importance of the long-range critical fluctuations.

\section{Summary and Discussion}
We have proposed the formulation to take into account the correction of the spatial correlations to the local self-energy obtained by DMFT.
It is demonstrated that the one-particle spectral intensity of the 2D Hubbard model at half-filling shows the pseudo-gap behavior in the central coherent quasiparticle peak due to the critical AF fluctuation.
It is important to note that the pseudo gap driven by the critical AF spin fluctuation is always accompanied by the enhancement of $1/T_1T$ in the NMR relaxation rate.
Thus, the spin-gap behavior\cite{Yasuoka89} in $1/T_1T$ of the hole-doped high-$T_c$ cuprates in the underdoped region requires alternative explanations such as the effect of critical superconducting fluctuations\cite{Yanase01,Kobayashi02}.
The specific heat is considerably enhanced by the AF short-range order, which is accompanied by the reduction of the double occupancy, indicating the enhanced tendency of the Mott localization.

Here we should note a deficiency of the approximate local dynamical susceptibility based on IPT.
The spectral intensity of the approximate dynamical spin susceptibility becomes slightly negative in high-energy region.
This is because the use of the energy-independent charge vertex in eq.~(\ref{ipt-susceptibility}).
Note that the charge vertex should become the bare $U$ in the high-energy limit.
This deficiency can be removed by the use of a charge vertex with proper energy dependence, which is evaluated by QMC for instance.
The QMC simulation with the Hirsch-Fye algorithm is of course more accurate than the present method, however, the computational cost is much more expensive, especially in lower temperatures and the strong $U$ region.
It should be emphasized that the deficiency of the present method is not so serious in the improved self-energy, since we consider only the nonlocal \textit{correction} of the susceptibilities, $\Delta\chi_\alpha(q)=\chi_\alpha(q)-\chi_\alpha(\epsilon_m)$, where $\Delta\chi_\alpha(q)$ vanishes in high-energy region as it should.
In other words, the essential local dynamics is sufficiently taken into account in the DMFT self-energy.

Throughout this work, we have updated the self-energy only \textit{once}.
Therefore, the spin fluctuation in our self-energy contains no feedback from the formation of the pseudo gap.
If such a feedback is taken into account, the critical AF fluctuation as well as the pseudo-gap behavior would slightly be suppressed.
In order to consider the feedback effect, one possible way is to solve the Dyson's equation in a self-consistent manner, keeping the local vertices and the DMFT self-energy unchanged.
This kind of self-consistent treatment includes multiple scattering from spin fluctuations, which would make the incoherent part of the single-particle spectrum much broader.

A further improvement of the present formulation should consider the local moment sum-rule (the fluctuation-dissipation theorem) for the spin susceptibility\cite{Vilk97}, which is slightly violated in the present formulation.
This sum rule becomes significantly important to discuss the anomalous temperature dependences in the vicinity of the quantum critical point\cite{Moriya95,Saso99,Millis93}.
In view of the success of the self-consistent renormalization theory\cite{Moriya95,Moriya85}, an introduction of a constant shift in the static component of the spin irreducible vertex may work out to satisfy the local-moment sum-rule.

On the physics of the 2D Hubbard model in the context of the high-$T_c$ cuprates, it is important to discuss the effect of an asymmetry of the tight-binding dispersion with longer-range hoppings, and of electron/hole doping\cite{Tremblay06}.
In region of finite dopings, it is also curious to see the superconducting instability based on the formula given in Appendix B.
They are left for the future work.

\section*{Acknowledgment}
The author would like to acknowledge stimulating discussions with T. Mutou, T. Takimoto and H. Yokoyama throughout this work. He also thank T. Saso, K. Miyake, Y. Kuramoto and A.-M. Tremblay for many valuable discussions.

\appendix
\section{Iterated Perturbation Theory}
In this Appendix, we summarize the formulation of the iterated perturbation theory (IPT) as a DMFT solver.
Motivated by the success of the second-order perturbation theory in the symmetric impurity Anderson model\cite{Yamada75}, the local self-energy in IPT is written in the interpolation form\cite{Georges92,Kajueter96,Kajueter96a,Potthoff97},
\begin{equation}\label{se-ipt}
\tilde{\Sigma}^{\rm L}(\omega_n)=\frac{A \Sigma^{(2)}(\omega_n)}{1-B\Sigma^{(2)}(\omega_n)},
\end{equation}
where the second-order self-energy is expressed as,
\begin{equation}
\Sigma^{(2)}(\omega_n)=U^2T\sum_{\epsilon_m}\bar{\chi}_0(\epsilon_m)\tilde{\cal G}(\epsilon_m+\omega_n),
\end{equation}
with $\bar{\chi}_0(\epsilon_m)=-T\sum_{\omega_n}\tilde{\cal G}(\omega_n)\tilde{\cal G}(\omega_n+\epsilon_m)$.
$\tilde{\cal G}(\omega_n)$ is the cavity Green's function or the ``bare'' Green's function in the fictitious impurity Anderson model, which contains all of the dynamical information of the other sites of the lattice.

The coefficients, $A$ and $B$, are determined so as to satisfy the high-energy and the atomic limits\cite{Kajueter96,Kajueter96a,Potthoff97}.
Their expressions are given by
\begin{subequations}
\begin{align}
&A=\frac{n(2-n)}{n_0(2-n_0)}, \\
&B=4\frac{\tilde{\mu}_f-\tilde{\mu}+U(1-n)}{U^2n_0(2-n_0)},
\end{align}
\end{subequations}
where the electron density $n=2T\sum_{\omega_n}G^{\rm L}(\omega_n)e^{i\omega_n0_+}$ and $n_0=2T\sum_{\omega_n}\tilde{\cal G}(\omega_n)e^{i\omega_n0_+}$.
Here the fictitious chemical potential of the mapped Anderson model, $\tilde{\mu}_f$, is introduced, which is determined by the condition in the low-energy limit\cite{Potthoff97a}.
Namely the Luttinger sum-rule is enforced to determine $\tilde{\mu}_f$ at $T=0$\cite{Kajueter96,Kajueter96a}.
For finite temperatures, a natural extension of the Luttinger sum-rule is the following identity\cite{Abrikosov63},
\begin{equation}
T\sum_{\omega_n}G^{\rm L}(\omega_n)\frac{\partial \Sigma^{\rm L}(\omega_n)}{\partial i\omega_n}=0,
\end{equation}
which is evaluated via the analytic continuation with the Pad\'e approximant.

Using the local self-energy, (\ref{se-ipt}), we obtain the local Green's function as
\begin{equation}
G^{\rm L}(\omega_n)=\frac{1}{N}\sum_{\mib{k}}G(k)=\int d\epsilon \frac{\rho_0(\epsilon)}{i\omega_n+\tilde{\mu}-\tilde{\Sigma}^{\rm L}(\omega_n)-\epsilon},
\label{G-local}
\end{equation}
where $\rho_0(\epsilon)$ is the bare DOS of the conduction electron.
The cavity Green's function is calculated by
\begin{equation}
1/\tilde{\cal G}(\omega_n)=1/G^{\rm L}(\omega_n)+\tilde{\mu}_f-\tilde{\mu}+\tilde{\Sigma}^{\rm L}(\omega_n).
\end{equation}
Then, the updated $\tilde{\cal G}(\omega_n)$ is used to calculate new self-energy until a convergence within the given accuracy is reached.
Note that in the symmetric case\cite{Georges92}, we have simply $\tilde{\mu}_f=\tilde{\mu}=0$ and $n=n_0=1$ ($A=1$ and $B=0$), and $\tilde{\Sigma}^{\rm L}(\omega_n)=\Sigma^{(2)}(\omega_n)$.

There is one technical comment in order to obtain an insulating solution.
The insulating solution has the form in the limit of $|\omega_n|\to0$\cite{Georges92,Rozenberg94},
\begin{subequations}
\begin{align}
&\tilde{\Sigma}^{\rm L}(\omega_n)\sim \frac{a}{i\omega_n}+\tilde{\Sigma}_{\rm inc}, \\
&G^{-1}(\omega_n)\sim \frac{i\omega_n}{b}-\frac{a}{i\omega_n}+R_{\rm inc}, \\
&\tilde{\cal G}^{-1}(\omega_n)\sim \frac{i\omega_n}{b}+S_{\rm inc}.
\end{align}
\end{subequations}
The last expression with the leading term $i\omega_n/b$ is obtained by the exact cancellation of the singular terms, $a/i\omega_n$ in $\tilde{\Sigma}^{\rm L}(\omega_n)$ and $G^{-1}(\omega_n)$.
In practice, however, numerical errors often leave the singular term $a/i\omega_n$ in $\tilde{\cal G}(\omega_n)$, which masks the leading term $i\omega_n/b$.
This results in the convergence to the metallic solution instead of the insulating one.
To avoid miscancellation, eq.~(\ref{G-local}) is modified to
\begin{equation}
G^{\rm L}(\omega_n)=\frac{1}{z_n}\left[1+\int d\epsilon \frac{\epsilon\rho_0(\epsilon)}{z_n-\epsilon}\right].
\end{equation}
with $z_n=i\omega_n+\tilde{\mu}-\tilde{\Sigma}^{\rm L}(\omega_n)$.
Then, the factor $1/z_n$ is responsible for the exact cancellation with $\tilde{\Sigma}^{\rm L}(\omega_n)$.

\section{Determination of Superconducting Transition Temperature}
Within the approximation proposed in this paper, the superconducting transition temperature, $T_c$, can also be determined.
In this appendix, we describe how to determine $T_c$ for the singlet and the triplet channels.

In general, $T_c$ is determined by the divergence of the static and homogeneous component of the renormalized vertex function, $\Gamma^*_\eta(k,k';0)$\cite{Abrikosov63}.
$\Gamma^*_\eta$ obeys the BS equation in the particle-particle (PP) channel,
\begin{multline}
\Gamma^*_\eta(k,k';0)=\Gamma_\eta^{\rm PP}(k,k';0)
\\
-\sum_{k''} \Gamma_\eta^{\rm PP}(k,k'';0)|G(k'')|^2\Gamma^*_\eta(k'',k';0),
\end{multline}
where $\Gamma^{\rm PP}_\eta$  ($\eta=\text{sin, tri}$) is the irreducible vertex in the singlet and the triplet PP channels.
In the vicinity of the transition, we can neglect the first term to yield
\begin{equation}
\Sigma_a(k)=-\lambda\sum_{k'} \Gamma^{\rm PP}_\eta(k,k';0)|G(k')|^2\Sigma_a(k'),
\end{equation}
where the anomalous self-energy is introduced as $\Sigma_a(k)=-\sum_{k'} \Gamma^*_\eta(k,k';0)\phi_\eta(k')$.
Then, $T_c$ is determined such that the maximum eigenvalue $\lambda$ of the above eigenvalue equation becomes unity.
In practice, the alternative expression of an eigenvalue problem for the anomalous propagator is useful,
\begin{equation}
\phi_\eta(k)=-\lambda|G(k)|^2\sum_{k'} \Gamma^{\rm PP}_\eta(k,k';0)\phi_\eta(k'),
\end{equation}
since $\phi_\eta(k)$ decays faster than $1/|\omega_n|^2$ in the high-energy limit.

\begin{figure}
\includegraphics[width=8.5cm]{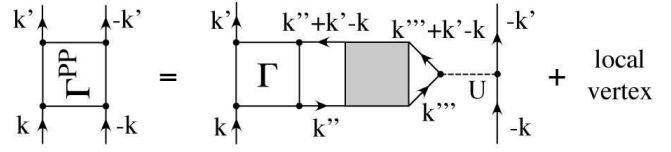}
\caption{The Feynman diagram of the particle-particle irreducible vertex expressed by the spin and the charge fluctuations.}
\label{pp-vertex}
\end{figure}

To solve the eigenvalue equation, it is necessary to give the explicit form of the irreducible vertices.
In general, they are given by the functional differentiation of the Luttinger-Ward functional with respect to the anomalous Green's function\cite{Allen01,Bickers89,Bickers91}, and they can be expressed in terms of the irreducible vertices in the particle-hole channel.
When the superconducting transition is triggered by the spin and/or the charge fluctuations, the dominant contribution to the irreducible vertices is given by the exchange of the single collective modes (so-called the Maki-Thompson process)\cite{Bickers89,Bickers91,Kyung03}.
Neglecting contributions other than the Maki-Thompson process, the irreducible vertices can be expressed within the present approximation as
\begin{subequations}
\begin{align}
&
\Gamma^{\rm PP}_{\rm sin}(k,k';0)=U-\frac{U}{2}\biggl[
\Gamma^{\rm L}_{\rm ch}(\omega_n'-\omega_n)\Delta\chi_{\rm ch}(k'-k)
\notag\\&\,\,
-3\Gamma^{\rm L}_{\rm sp}(\omega_n'-\omega_n)\Delta\chi_{\rm sp}(k'-k)\biggr]_{\rm even} + \Gamma^{\rm L}_{\rm sin}(\omega_n,\omega_n';0),\\
&
\Gamma^{\rm PP}_{\rm tri}(k,k';0)=-\frac{U}{2}\biggl[
\Gamma^{\rm L}_{\rm ch}(\omega_n'-\omega_n)\Delta\chi_{\rm ch}(k'-k)
\notag\\&\quad\quad\quad\quad
+\Gamma^{\rm L}_{\rm sp}(\omega_n'-\omega_n)\Delta\chi_{\rm sp}(k'-k)\biggr]_{\rm odd},
\end{align}
\end{subequations}
where the subscript even (odd) indicates the even (odd) part with respect to $k\to -k$.
Here, $\Gamma^{\rm L}_{\rm sin}$ is associated with the local anomalous self-energy, $\tilde{\Sigma}^{\rm L}_a(\omega_n)$.
Its explicit form is not necessary in considering the anisotropic pairings, since $\Gamma^{\rm L}_{\rm sin}$ does not affect the transition temperature for anisotropic pairings.
The Feynman diagram of the PP irreducible vertex is shown in Fig.~\ref{pp-vertex}.


\begin{thebibliography}{99}
\bibitem{Potthoff97a} In addition to the Luttinger sum-rule, possible conditions are examined in ref.~\citen{Potthoff97}.
\end{thebibliography}
\end{document}